# Optical Vortex Spin-Orbit Control of Refractive Index in Iron Garnets

**Seth Nelson[1], Cong Yu[1], Daniel Watson[1], Shahrzad Ramtinfard[1], and Miguel Levy[1,*]**

[1]*Physics Department, Michigan Technological University, Houghton, Michigan USA*
*mlevy@mtu.edu

**Abstract:** The interaction between light's angular momentum (AM) and material systems has unlocked new avenues in structured photonics, including in magneto-optical (MO) materials. While spin angular momentum (SAM) effects in MO systems are well-established, orbital angular momentum (OAM) introduces novel opportunities for new nonreciprocal light-matter interactions. In this study, we demonstrate a unique optical phenomenon where OAM states undergo state-specific nonreciprocal operation within an MO medium, reducing Faraday rotation. This effect arises from transverse momentum transfer into the material, inducing spin-orbit coupling (SOC) at a perturbed electronic transition rate. The resulting OAM-dependent optical SOC modifies the material's refractive index, directly linking structured light and MO response. Our findings extend previous observations of paraxial beams and reveal a deeper fundamental mechanism governing OAM-driven nonreciprocal interactions. These insights pave the way for OAM-selective nonreciprocal photonic devices, chiral optical logic, quantum memory elements, and ultrafast spintronic architectures. This work advances MO integration with structured light for enhanced control over photonic and spintronic systems.

## 1. Introduction

The interplay between optical angular momentum (AM) and material systems has been a cornerstone in modern photonics. This subject has opened up new frontiers in light-matter interactions, particularly in structured and topological optics [1,2]. Spin angular momentum (SAM) has been extensively studied as it is a fundamental photon property, with left- and right-circular polarizations corresponding to photons having either $\pm 1\hbar$ SAM. Helical phase fronts with topological charge characterize light's orbital angular momentum (OAM). This is most commonly demonstrated with Laguerre-Gaussian vortex beams where each photon within the beam contains $\pm l\hbar$ OAM, for integer rotational mode numbers l [3-5]. The ability to manipulate AM has enabled diverse applications in optical communications, quantum information processing, and chiral light-matter interactions [3-6].

The transfer of AM between SAM and OAM is central to the control and modulation of structured light [7-9]. This transfer occurs in specially designed anisotropic and inhomogeneous media, where birefringent elements or geometric phase optics induce spin-to-orbital conversion [10-12]. Recent studies have demonstrated that structured optical beams carrying OAM can induce spin-to-orbital coupling (SOC) in various optical systems, enabling new ways to manipulate light-matter interactions [7,13,14]. In particular, inhomogeneous and anisotropic media have been shown to facilitate SAM-to-OAM conversion, generating optical vortices and chiral wavefronts [6,15-17]. These effects strongly impact nonreciprocal photonic devices, topological photonics, and magneto-optically active materials, where light's angular momentum interacts with electronic and magnetic degrees of freedom [18-20].

Magneto-optical (MO) materials, known for their anisotropic and nonreciprocal properties, have long been studied for spin angular momentum (SAM) interactions. Recently, interest has evolved to encompass their coupling with orbital angular momentum (OAM), opening new avenues for photonic and magnetophotonic applications [21]. These materials' nonreciprocal response mimic time-reversal symmetry breaking, enabling OAM-driven effects such as all-optical magnetization switching (AOS), where SAM and OAM influence ultrafast spin dynamics in ferromagnetic systems [13,22,23]. Beyond magnetization control, OAM transfers momentum to magnetic systems, impacting quasiparticles like magnons and skyrmions. Optical vortex beams can exert torque on magnetic textures, inducing skyrmion rotation and enabling magnetic tweezers [13,19,24,25]. Experimental demonstrations of optical spin–orbit torque further establish MO materials as a bridge between photonics and spintronics [24-27].

Recent discoveries in structured light–MO interactions reveal topological magneto-optical effects, including OAM-selective modulation and intensity-dependent Faraday rotation [28]. SAM phase shifts vary uniquely across OAM states in paraxial Laguerre-Gaussian beams, experimentally manifesting as reduced Faraday rotation while preserving beam topology [29]. These effects introduce a new degree of freedom for chiral photonic information processing, advancing the integration of structured light with MO systems.

In the work reported herewith, we demonstrate that OAM states undergo nonreciprocal state-specific operation while propagating through MO materials. This operation is observed to result in a reduction in Faraday rotation. We show that this reduction is observed in collimated Laguerre-Gaussian beams, without the unique Gouy phases for OAM states previously studied in strongly focused paraxial beams [29]. Instead, we reveal transverse momentum transfer into the MO material due to the propagating OAM states. This transfer of momentum results in the optical SOC of the SAM states, inducing electronic excitations at a perturbed rate. The electronic perturbation generates OAM state-specific optical SOC and refractive index modification, sequentially changing the material-specific Faraday rotation. The treatment reveals the electronic origins of these effects and provides evidence of OAM-induced skyrmionics phenomena in magneto-optic systems. These findings impact the creation of nonreciprocal OAM state-specific operation, providing a new framework for designing OAM-driven nonreciprocal photonic devices, chiral optical logic, chiral quantum memory elements, and ultrafast spintronic architectures. Thus, they advance the integration of magneto-optics with structured light and contribute to achieve better control over light-matter interactions.

## 2. Optical OAM Perturbation to the Hamiltonian

Here we demonstrate that this momentum transfer leads to momentum-dependent perturbations of the electronic states in MO materials. We take a fundamental approach to this analysis by finding the perturbation in the electronic orbital Hamiltonian, the change in virtual electronic excitations and consequentially the change in refractive index. Bismuth-substituted iron garnets (BiIG) are used for experimental verification in this work, and BiIG parameters for our theoretical calculations. These materials have long been used for optical isolation and other photonic applications [30-33].

We begin with the standard Hamiltonian formulation, shown as Eq. 1, describing the motion of a charge $e^-$ in an EM field [34-36].

$$\hat{H} = \frac{1}{2m_e}\left(\hat{p} + \frac{e}{c}\vec{A}\right)^2 + V \qquad \text{Eq. 1}$$

Here $\hat{p}$ is the momentum operator, while $\vec{A}$ is the vector potential for the propagating OAM beam through the MO material. $V$ is the MO crystal field potential, which is discussed extensively in [36].

Since MO materials primarily exhibit linear optical properties, the last term in the quadratic expansion, $\frac{e^2 A^2}{2m_e c^2}$, will be neglected. The terms $\frac{\hat{p}\cdot\hat{p}}{2m_e} + V$ correspond to the unperturbed material Hamiltonian, whereas $\frac{e}{2m_e c}\hat{p}\cdot\vec{A}$ will be further expanded when applied to the quantum state the Hamiltonian is operating on. This expansion results in an additional term, describing the OAM beam being operated on by the momentum operator. Typically, for non-OAM-carrying beams, this term is zero, as the momentum is purely translational in the direction of propagation. However, for an OAM carrying beam, such as a Laguerre-Gaussian beam, it does not vanish. Hence, we must include the perturbation described in Eq. 2 for the MO Hamiltonian.

$$\hat{H}_p = \frac{e}{m_e c}\vec{A}\cdot\hat{p} + \frac{e}{2m_e c}(\hat{p}\cdot\vec{A}) \qquad \text{Eq. 2}$$

Based on the gauge chosen for the vector potential, we rewrite $\frac{e}{m_e c}\vec{A}\cdot\hat{p}$ as $-e\,\vec{r}\cdot\vec{E}$ [34]. This is the traditional dipolar perturbation of the Hamiltonian that yields the unique refractive indices for left and right polarizations [36]. The results of which have been extensively studied and examined. However, due to the presence of the OAM beam, there will be an additional term, shown as Eq. 3, which will add a new perturbative effect to the MO material electronic transitions.

$$\hat{H}_{p,OAM} = \frac{e}{2m_e c}(\hat{p}\cdot\vec{A}) \qquad \text{Eq. 3}$$

To demonstrate this, let us define the vector potential as Eq. 4 for the propagating OAM Laguerre-Gaussian beam [37].

$$\vec{A}(\rho,\phi,z) = \widehat{e_\sigma}\frac{A_0\omega_0}{\omega(z)}\left(\frac{\sqrt{2}\rho}{\omega(z)}\right)^{|\ell|} L_p^{|\ell|}\left(\frac{2\rho^2}{\omega(z)^2}\right) e^{-\frac{\rho^2}{\omega(z)^2}} e^{ik\left(z+\frac{\rho^2}{2R(z)}\right)} e^{i((\ell\pm 1)\phi + \phi_G(z))} \qquad \text{Eq. 4}$$

In the above expression, the Rayleigh length $z_R = \frac{\pi \omega_0^2}{\lambda}$, the beam radius $\omega(z) = \omega_0\sqrt{1+\left(\frac{z}{z_R}\right)^2}$, the radius of the beam curvature $R(z) = z + \frac{z_R^2}{z}$, and the Gouy Phase $\phi_G(z) = -(|\ell| + 2p + 1)\tan^{-1}\left(\frac{z}{z_R}\right)$. The primary polarization states will be circular polarizations when propagating within MO material. These are defined as $\widehat{e_\sigma} = \hat{\rho} \pm i\hat{\phi}$. Following the compilation of these equations into the OAM perturbed Hamiltonian, we arrive at the simplified perturbed Hamiltonian, in Eq. 5. The details of this Hamiltonian are further examined in the Supplement [38].

$$\hat{H}_{p,OAM} = \frac{-i\hbar e}{2m_e c}\frac{A_0\omega_0}{\omega(z)}\left(\frac{\sqrt{2}\rho}{\omega(z)}\right)^{|\ell|} \left(\frac{ik\rho}{R(z)}L_p^{|\ell|}\left(\frac{2\rho^2}{\omega(z)^2}\right) + \frac{|\ell|\mp\ell}{\rho}L_p^{|\ell|}\left(\frac{2\rho^2}{\omega(z)^2}\right) - \frac{2\rho}{\omega(z)^2}L_p^{|\ell|}\left(\frac{2\rho^2}{\omega(z)^2}\right) - \frac{4\rho}{\omega(z)}L_{p-1}^{|\ell|+1}\left(\frac{2\rho^2}{\omega(z)^2}\right)\right) e^{-\frac{\rho^2}{\omega(z)^2}} e^{ik\left(z+\frac{\rho^2}{2R(z)}\right)} e^{i((\ell\pm 1)\phi + \phi_G(z))} \qquad \text{Eq. 5}$$

We now make some physically motivated substitutions when examining this perturbation at the point of a lattice within the MO material, discussed in the Supplement. This allows us to rewrite the perturbed Hamiltonian as Eq. 6 where $A(\rho_0, \phi_0, z_0)$, shown in Eq. 7, is the constant amplitude at the given lattice point.

$$\hat{H}_{p,OAM} = A(\rho_0, \phi_0, z_0) \left( \frac{(|\ell|\mp\ell)}{2\rho_0} - \frac{2\rho_0}{2\omega_c^2} - \frac{4\rho_0}{2\omega_c} \frac{L_{p-1}^{|\ell|+1}\left(\frac{2\rho_0^2}{\omega_c^2}\right)}{L_p^{|\ell|}\left(\frac{2\rho_0^2}{\omega_c^2}\right)} \right) e^{\pm i\phi'} \qquad \text{Eq. 6}$$

$$A(\rho_0, \phi_0, z_0) = \frac{-i\hbar e}{m_e c} \frac{A_0 \omega_0}{\omega_c} \left( \frac{\sqrt{2}\rho_0}{\omega_c} \right)^{|\ell|} L_p^{|\ell|}\left(\frac{2\rho_0^2}{\omega_c^2}\right) e^{-\frac{\rho_0^2}{\omega_c^2}} e^{i(kz_0 + \ell\phi_0 + \phi_G(z_0))} \qquad \text{Eq. 7}$$

Notice that the perturbation contains the primary term $|\ell| \mp \ell$. This will result in optical SOC, while the SAM has different refractive indices within the MO material, creating a nonreciprocal operation on the optical OAM states.

## 3. Perturbation of the MO Refractive Index

We use Fermi's golden rule and the Kramers-Kronig relation to transition from the perturbation Hamiltonian to the refractive index perturbation. As shown in [36], the transitions seen in bismuth-substituted iron garnets are $^6S \to {}^4P$ and $^3P$ Hybrids. Hence the virtual excitations will take a ground state $|0\rangle$ to the excited $|\pm 1\rangle$ states with $\Delta L_z = \pm 1$. These perturbations of the excitation probabilities allow us to find the perturbation of the MO refractive index due to the transfer of transverse momentum from the OAM Laguerre-Gaussian beam. The net effect is a modification of the rate of electronic transition, shown in Eq. 8, due to the propagating OAM Laguerre-Gaussian beam [35].

$$\alpha(\omega) = \frac{2\pi}{\hbar} |\langle 1|\hat{H}_{p,OAM}|0\rangle|^2 \delta(E_1 - E_0 - \hbar\omega) \qquad \text{Eq. 8}$$

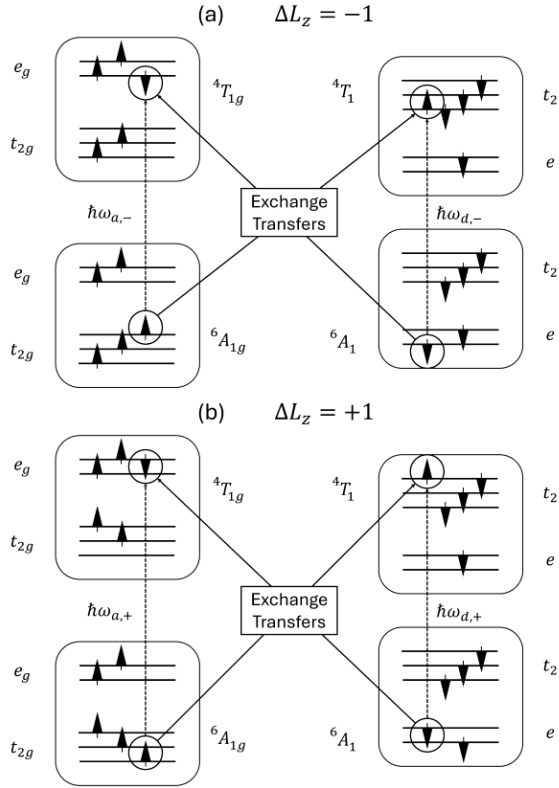

*Fig. 1: Schematic diagrams for both photonics SAM of the intersublattice pair transfer for the super-exchange mechanism. Due to the spin-orbit coupling, (a) corresponds to $+1\hbar$ SAM, and (b) corresponds to $-1\hbar$ SAM [36].*

The electrons will be super-exchanged for the two different sublattices within the BiIG lattice. This entails the following chemical transfer processes [36], shown in detail in Fig. 1:

$$Fe_a^{3+}\left(t_{2g}^3 e_g^2\right) + Fe_d^{3+}\left(t_g^3 e^2\right) \to Fe_a^{4+}\left(t_{2g}^2 e_g^2\right) + Fe_d^{2+}\left(t_g^4 e^2\right) - \hbar\omega_a - \Delta(IP)$$

$$Fe_a^{4+}\left(t_{2g}^2 e_g^2\right) + Fe_d^{2+}\left(t_g^4 e^2\right) \to Fe_a^{3+}\left(t_{2g}^2 e_g^3\right) + Fe_d^{3+}\left(t_g^4 e^1\right) - (\hbar\omega_a + E_{ex}) + \Delta(IP)$$

Here $\hbar\omega_a$ and $\hbar\omega_d$ correspond to the energy of the virtual photons inducing the excitations, $\Delta(IP)$ is the ionization potential energy, and $E_{ex}$ being the stabilization energy created by the super-exchange mechanism. Hence $E_{ex} = \hbar\omega_d$, so that in total the process becomes:

$$Fe_a^{3+}\left(A_{1g}\right) + Fe_d^{3+}(A_1) \to Fe_a^{3+}\left(T_{1g}\right) + Fe_d^{3+}(T_1) - (\hbar\omega_a + \hbar\omega_d)$$

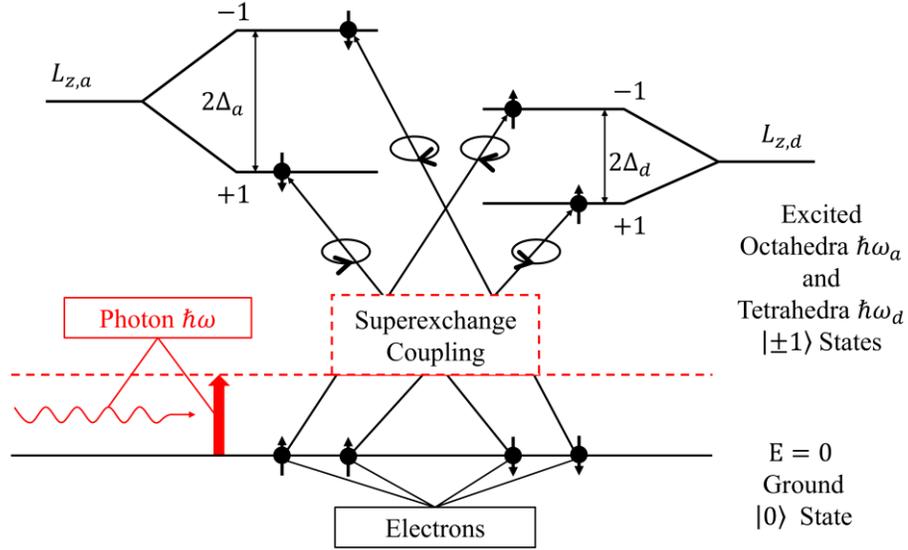

*Fig. 2: Virtual state excitation of the electric-dipole transitions in the presence of a photon with the energy $\omega \neq \omega_a \neq \omega_d$ [36].*

Virtual photons of energy $\hbar\omega_a$ or $\hbar\omega_d$ will cause both $T_{1g}$ and $T_1$ excitation due to the super-exchange mechanism. A complete picture of this virtual excitation is shown in Fig. 2. When a photon of a given frequency $\omega \neq \omega_a \neq \omega_d$ enters the material, it induces virtual transitions, slightly changing the probability of electronic transition. The rate of virtual electronic transition is thus physically modified to the actual energy levels via $\hbar\omega_a$ or $\hbar\omega_d$ within the MO lattice. From this modified transition rate, one can calculate the perturbation to the MO material's refractive index using the Kramers-Kronig rule [39]. The primary relationship is given by Eq. 9.

$$n_{p,\pm}(\omega)^2 = \frac{2}{\pi}\int_0^\infty \frac{\omega' \alpha_\pm(\omega)}{\omega'^2 - \omega^2} d\omega' \qquad \text{Eq. 9}$$

Where $\alpha_\pm(\omega)$ is the modified rate of electronic transition; after evaluating the rate of transitions, we arrive at the change in the refractive index due to the perturbed Hamiltonian at the given lattice point. However, to calculate the effective refractive index of the entire Laguerre-Gaussian beam, we must sum over all lattice points. To normalize this calculation, we divide by the total energy of the Laguerre-Gaussian beam. The end result is a general perturbation of the MO refractive index due to the OAM beam, given by Eq. 10.

$$n_{p,\pm}(\omega) = \mp \sqrt{\left(\frac{2\omega_a}{(\omega_a^2-\omega^2)} + \frac{2\omega_d}{(\omega_d^2-\omega^2)}\right)\frac{\Sigma_{\rho_0,\phi_0,z_0}|\langle\pm 1|\hat{H}_{p,OAM}(\rho_0,\phi_0,z_0)|0\rangle|^2}{\hbar 4\pi\varepsilon_0 \int|\vec{E}|^2 dV}} \qquad \text{Eq. 10}$$

This refractive index perturbation is produced by the nonreciprocal optical SOC. There will be a nonreciprocal reduction in Faraday rotation depending on the sign of the spin and orbital state. Explicitly, it takes the form $n_{eff,\pm} = n_\pm \mp n_{p,\ell}$. For linearly polarized light, this perturbation to the refractive index will reduce the Faraday rotation as shown in Eq. 11.

$$\theta_{F,\ell} = \theta_F - \frac{\pi}{\lambda} n_{p,\ell} d \qquad \text{Eq. 11}$$

Crucially, this reduction in Faraday rotation is nonreciprocal and specific to each OAM state. Thus, sending the Laguerre-Gaussian beam back into the MO material will not return the original linear polarization or the typical Faraday rotation but a unique linear polarization rotation for each OAM state. Of course, this will depend on the propagation direction regarding the MO material magnetization [40, 41], enabling nonreciprocal and controllable MO magnetization operation on the individual OAM states.

## 4. Experimental Methods and Results

Recently, it has been demonstrated that minimal changes in the refractive index can be precisely measured using OAM interferometry [42, 43]. We implemented these methods to experimentally verify our proposed nonreciprocal phenomenon to create an optical OAM-based Mach-Zehnder interferometer. Figure 3 shows the experimental setup we used for this purpose. A flower petal pattern results from the interference of opposite topologically charged beams, as described in the Methods section. The presence of a magnetically poled bismuth-substituted latching iron garnet produces a rotation of the petal pattern, yielding a precise measure of the refractive index of the material for right- and left-circularly polarized OAM light.

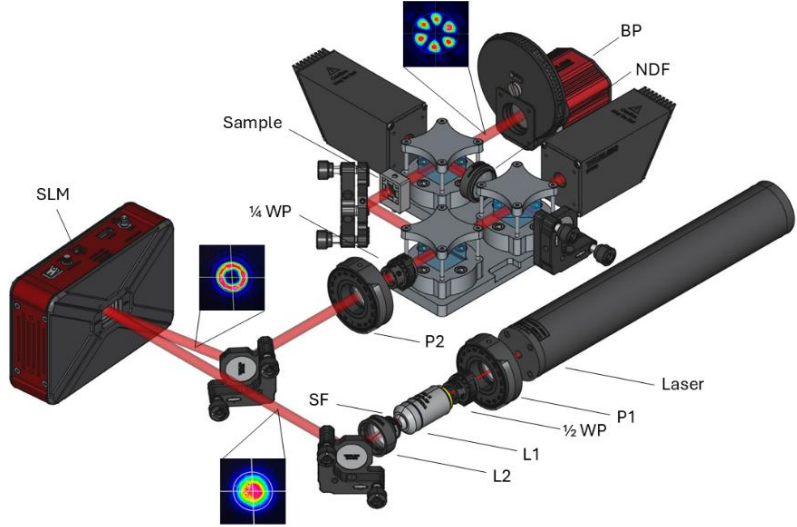

Fig. 3: Full experimental diagram demonstrating the measurement of the OAM state specific phases, and thus refractive indices, after passing through the target sample

For our experimental analysis, we used a 632 nm HeNe laser beam, expanded and collimated, before reflecting it off a spatial light modulator (SLM) to generate OAM states. Following the generation of OAM states, the beam was passed through a polarizer and a quarter wave plate.

The quarter-wave plate was mounted onto a rotational optics mount, allowing for the adjustment of polarization, including linear and circular polarizations. The beam was then directed into a Mach-Zehnder interferometer, where one arm passed through the magnetically poled bismuth-substituted latching iron garnet sample, which underwent optical AM-specific operation within the material. In contrast, the other arm inverted the OAM topological charge and thus served as a reference. A 1.5 neutral density filter was used to reduce the reference intensity to match the sample arm's intensity level. Upon recombination at a beam splitter, the interference pattern formed a characteristic petal-like structure whose angular displacement correlated directly with the refractive index of the MO sample. A beam profiling camera captured images of the petal interference pattern generated. By rotating the quarter-wave plate, the input OAM beam's polarization was adjusted, thereby changing the effective refractive index of the OAM beam. This change in the effective refractive index would create a phase difference to the reference OAM beam. Such that rotating the quarter waveplate would cause the petal interference pattern to rotate, providing a direct measurement of the refractive index variations. The whole picture of this experiment is shown in Fig. 3.

To equate the observed petal rotation with the change in refractive index, as shown in Eq. 12, we employed the same mathematical methods previously demonstrated for interferometry refractive index measurements using OAM beams [42, 43].

$$\Delta n = \frac{\lambda \ell \, \theta_{rot}}{\pi d} \qquad \text{Eq. 12}$$

Where $\theta_{rot}$ is the petal interference pattern's measured rotation and d is the MO sample's thickness. This change in refractive indices for each OAM state and SAM state can be directly substituted into Eq. 11 to calculate the reduction in experimentally measured Faraday rotation. To measure the petal rotation, we converted the images acquired by the beam profiler to grayscale and assigned each pixel a value from 0 to 1, depending on its grayscale color. This allowed us to create a pixel matrix for each beam profiler image for pixel-to-pixel comparison of the different polarizations for each OAM state. We then applied a rotation matrix to one of the images being compared and determined the angle at which the subtraction of the rotated matrix would result in a null matrix. We analyzed left- and right-circular polarizations, as well as vertical and horizontal polarizations. Allowing for the direct determination of the difference in refractive indices for the circular polarizations. This allows for extracting the OAM state perturbation from the MO material refractive index following normalization to the measured Faraday rotation for non-OAM beams. Linear polarizations allow for the normalization of these measurements. We obtained all the measurements described to observe the nonreciprocal operation on the Laguerre-Gaussian beam, with the beam propagating parallel and anti-parallel to the material internal magnetization. A detailed diagram of our experiment is shown in Fig. 3.

## 5. Discussion

To compare the experimental results and the theoretical predictions, we numerically solved the perturbation to the refractive index for the bismuth-substituted iron garnets for OAM states with topological charges ranging from 0 to 10. All parameters used in the numerical solution were experimentally obtained for OAM beams of 632 nm wavelength. The perturbation to the SAM specific refractive index was found for both positive and negative OAM topological charge. Due to the nonreciprocal nature of the SOC phenomena described in the theory section, the SAM states will share the same refractive index perturbation. This was numerically verified for all OAM states. Still, this perturbation only occurs for one SAM state depending on the sign of the OAM topological charge. Perturbation to the refractive index is an increase or decrease in value depending on the sign of the SAM and OAM. This results in a nonreciprocal operation on the OAM state-specific polarization, reducing Faraday rotation. Using the experimental methods described, we verified these predictions for all permutations of SAM and OAM signs. We found that the difference between the material-specific refractive indices for circular polarizations decreased. This will result in a reduction of the Faraday rotation dependent on the Laguerre-Gaussian OAM state. We then compared the experimental and theoretical total reduction in Faraday rotation, shown in Fig. 4.

Within Fig. 4, the theoretical predictions and experimental results are closely related. This is especially true for OAM states with a topological charge less than 7. In fact, these results closely align with the previously reported Faraday rotation reduction for OAM states in the paraxial approximation [29]. Imperatively, we have demonstrated these results for a nonparaxial Laguerre-Gaussian beam, which means that the phenomenon behind the reduction in Faraday rotation for OAM states we predicted and observed is not due to the Gouy phase. We computationally verify this assertion via COMSOL. But, instead, it is due to the transfer of OAM into the material and the subsequent SOC. Fundamentally changing the material response to the OAM states. However, as the topological charge increases, the experimental results and theoretical predictions slightly diverge. This is due to the increase in a global phase for the Laguerre-Gaussian beam after propagating through the bismuth-substituted iron garnet sample. This global phase (observed for all polarizations, circular and linear) increase depended on the time lapse the Laguerre-Gaussian beam was passing through the sample. Similar optical phase increases have been previously reported in the case of spin orbit torque (SOT), where there is a change in AM in time [44-46], suggesting that optical SOT operates on the bismuth-substituted iron garnet sample. Although the theoretical construct described here does not explicitly account for time dependence, our proposed theory predicts that there must be a transfer of AM to the sample. This transfer in AM would, of course, depend on the intensity of the Laguerre-Gaussian beam and its exposure time.

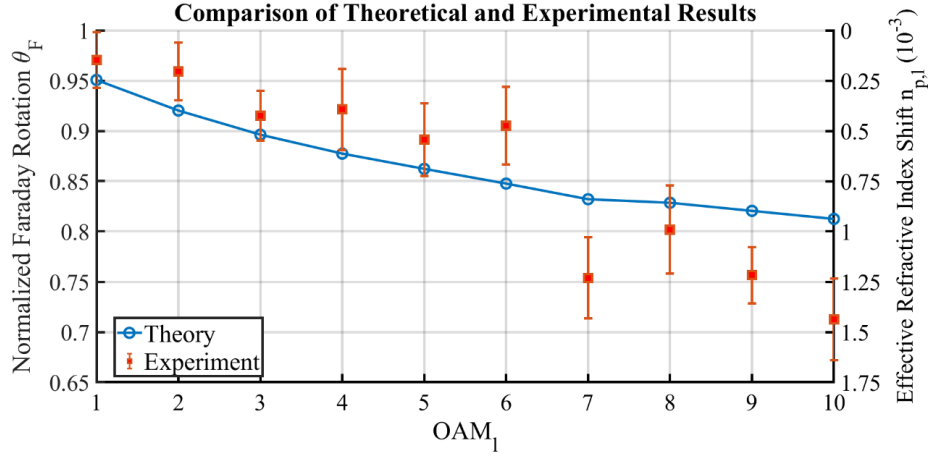

Fig. 4: Comparison of predicted theoretical reduction in Faraday rotation and experimentally measured and normalized to the measured traditional Faraday rotation for a simple Gaussian beam carrying no topological charge, no OAM.

We took our experimental measurements in approximately 10-15 second intervals to account for this time dependency. Following the measurement, we blocked the Laguerre-Gaussian beam from entering the sample. This reduced the time exposure and increased the accuracy of the measurements, particularly for smaller topological charges. However, with an increase in topological charge we observed an increase in the acquisition of this global phase. Resulting in less accurate experimental measurements for higher topological charges. Further exploration of this new phenomenon should reveal the extent of the time dependency and optical SOT and be prioritized for future research.

Our work here demonstrates a new optical phenomenon in which OAM states undergo nonreciprocal state-specific operation while propagating through MO material. This operation is experimentally observed as a reduction in Faraday rotation. Importantly, this reduction was demonstrated for collimated Laguerre Gaussian beams without the unique Gouy phases for the different OAM states in strongly focused paraxial beams [29]. Instead, the OAM states introduced a perturbation in the MO material electronic transitions and orbitals. We created OAM state-specific optical spin-orbit coupling, refractive indices, and sequentially OAM state-specific Faraday rotation. Our results here aligned with those previously seen for paraxial beams [29], however, we expanded the results to a more fundamental level. Demonstrating that the paraxial approximation is not required for the OAM state-specific Faraday rotation reduction. This means that lenses are not required for the phenomena we found, significantly decreasing the complexity of the Faraday rotation reduction. Importantly, we demonstrated that this phenomenon is nonreciprocal. Enabling the development of nonreciprocal operations tailored to specific OAM states, this approach introduces a new framework for designing OAM-based nonreciprocal photonic devices, chiral optical logic systems, chiral quantum memory components, and ultrafast spintronic architectures.

## 6. Conclusion

We present a new optical OAM perturbation to the electronic transition model and subsequent refractive index of bismuth-substituted iron garnets. We demonstrate that this perturbation yields nonreciprocal operation on OAM state-specific polarizations in magneto-optical materials. We experimentally verify our predicted phenomena using OAM-based interferometry. There is a strong agreement between our proposed theory and experimental findings, particularly for lower topological charges. Furthermore, we observed a time-dependent global phase accumulation for the OAM states passing through the bismuth-substituted iron garnets. This suggests the influence of optical spin-orbit torque (SOT), warranting further investigation. These findings open new possibilities for OAM-driven nonreciprocal photonic devices, including isolators, modulators, and polarization controllers, advancing chiral optical logic, quantum memory elements, and ultrafast spintronic architectures.

### Acknowledgements

M. Levy and S. Nelson gratefully acknowledge support from Photonica, Inc. and the Henes Center for Quantum Phenomena. S. Nelson gratefully acknowledges support from the Michigan Technological University Finishing Fellowship. M. Levy and S. Nelson gratefully acknowledge the use and assistance of lab equipment borrowed from Dr. Christopher Middlebrook and Micheal Maurer.

### Disclosures

The authors declare no conflicts of interest.

### Data Availability

Data underlying the results presented in this paper are not publicly available at this time but may be obtained from the authors upon reasonable request.